\documentclass[preprint,
amsmath,amssymb,amsfonts,aps,prd,superscriptaddress,nofootinbib]{revtex4-1}
\usepackage{ifsym}
\usepackage{MnSymbol}
\usepackage[usenames,dvipsnames]{xcolor}
\usepackage{hyperref}
\hypersetup{
    colorlinks,
    urlcolor=Maroon,
    linkcolor=BlueViolet,
    citecolor=Bittersweet
}

\def\be{\begin{equation}}
\def\ee{\end{equation}}
\def\ba{\begin{eqnarray}}
\def\ea{\end{eqnarray}}
\def\nl{\nonumber\\}

\def\CP1{\mathbb{CP}^1}
\def\SL2C{\mathrm{SL}(2,\mathbb{C})}

\def\Z2{\mathbb{Z}_2}

\def\su2{{SU(2)}}

\def\[{\left[}
\def\]{\right]}

\def\({\left(}
\def\){\right)}
\def\[{\left[}
\def\]{\right]}

\def\<{\langle}
\def\>{\rangle}

\def\i2{\frac{i}{2}}

\def\2F1{\,_2{\rm F}_1}

\newcommand{\beq}{\begin{equation}}
\newcommand{\eeq}{\end{equation}}
\newcommand{\beqq}{\begin{equation*}}
\newcommand{\eeqq}{\end{equation*}}
\newcommand\beqa{\begin{eqnarray}}
\newcommand\eeqa{\end{eqnarray}}
\newcommand\beqaa{\begin{eqnarray*}}
\newcommand\eeqaa{\end{eqnarray*}}
\newcommand\bea{\begin{array}}
\newcommand\eea{\end{array}}

\begin{document}


\title{New Double Soft Emission Theorems}


\author{Freddy Cachazo}
\email{fcachazo@perimeterinstitute.ca}
\affiliation{Perimeter Institute for Theoretical Physics, Waterloo, ON N2L 2Y5, Canada}
\author{Song He}
\email{she@perimeterinstitute.ca}
\affiliation{Perimeter Institute for Theoretical Physics, Waterloo, ON N2L 2Y5, Canada}
\affiliation{School of Natural Sciences, Institute for Advanced Study, Princeton, NJ 08540, USA}
\author{Ellis Ye Yuan}
\email{yyuan@perimeterinstitute.ca}
\affiliation{Perimeter Institute for Theoretical Physics, Waterloo, ON N2L 2Y5, Canada}
\affiliation{Department of Physics \& Astronomy, University of Waterloo, Waterloo, ON N2L 3G1, Canada}


\date{\today}

\begin{abstract}

We study the behavior of the tree-level S-matrix of a variety of theories as two particles become soft. By analogy with the recently found subleading soft theorems for gravitons and gluons, we explore subleading terms in double soft emissions. We first consider double soft scalar emissions and find subleading terms that are controlled by the angular momentum operator acting on hard particles. The order of the subleading theorems depends on the presence or not of color structures. Next we obtain a compact formula for the leading term in a double soft photon emission. The theories studied are a special Galileon, DBI, Einstein--Maxwell--Scalar, NLSM and Yang--Mills--Scalar. We use the recently found CHY representation of these theories in order to give a simple proof of the leading order part of all these theorems.

\end{abstract}


\maketitle

\section{Introduction and Summary of Results}\label{sec:intro}

In 2014 subleading soft factors were found for the single soft emission of gravitons in Einstein gravity \cite{Cachazo:2014fwa}. The structure of a $(n+1)$-graviton amplitude in the limit when $k_{n+1}$ is soft is given by
\be\label{gravity}
{\cal M}_{n+1} = (S^{(0)}_{\rm gravity} +S^{(1)}_{\rm gravity}+S^{(2)}_{\rm gravity} ){\cal M}_{n} + {\cal O}(\tau^2 )
\ee
where $\tau$ is a parameter that controls the soft limit, $k_{n+1}^\mu = \tau q^\mu$, and
\be
S^{(0)}_{\rm gravity} = \sum_{a=1}^n\frac{\epsilon_{\mu\nu}k^\mu_a\, k^\nu_a}{k_{n+1}\cdot k_a}
\ee
is Weinberg's universal soft factor \cite{Weinberg:1965nx}. In this formula $\epsilon_{\mu\nu}$ is the polarization tensor of the soft particle. The next two soft factors were found by Strominger and one of the authors in \cite{Cachazo:2014fwa} and are given by
\be
S^{(1)}_{\rm gravity} = \sum_{a=1}^n\frac{\epsilon_{\mu\nu}k^\mu_a (k_{n+1,\rho}J^{\rho\nu}_a)}{k_{n+1}\cdot k_a}, \quad S^{(2)}_{\rm gravity} = \frac{1}{2}\sum_{a=1}^n\frac{\epsilon_{\mu\nu}(k_{n+1,\rho}J^{\rho\mu}_a) (k_{n+1,\sigma}J^{\sigma\nu}_a)}{k_{n+1}\cdot k_a}.
\ee
In these formulas $J^{\mu\nu}_a$ is proportional to the total angular momentum operator acting on the $a^{\rm th}$ particle. For example, when $a$ is a scalar particle
\be
J^{\mu\nu}_{a,\text{scalar}} \equiv k_a^\mu\frac{\partial}{\partial k_{a,\nu}} - k_a^\nu\frac{\partial}{\partial k_{a,\mu}}.
\ee
By analogy with the gravity construction, Casali identified a sub-leading factor for the emission of a soft gluon in Yang--Mills \cite{Casali:2014xpa}. For a $U(N)$ color-ordered partial amplitude the structure becomes
\be\label{gauge}
{\cal M}(1,2,\ldots ,n,n+1) = (S^{(0)}_{\rm YM} + S^{(1)}_{\rm YM}){\cal M}(1,2,\ldots ,n) + {\cal O}(\tau ),
\ee
where
\be
S^{(0)}_{\rm YM} = \frac{\epsilon\cdot k_n}{k_{n+1}\cdot k_n} - \frac{\epsilon\cdot k_1}{k_{n+1}\cdot k_1}, \qquad S^{(1)}_{\rm YM} =\frac{\epsilon_{\mu}(k_{n+1,\rho}J_n^{\mu\rho})}{k_{n+1}\cdot k_n}-\frac{\epsilon_{\mu}(k_{n+1,\rho}J_1^{\mu\rho})}{k_{n+1}\cdot k_1}.
\ee
Sub-leading soft theorems have a long history dating back to the 50's (see e.g.,~\cite{Low:1958sn, *Burnett:1967km, Gross:1968in, *Jackiw:1968zza, Laenen:2008gt, *Laenen:2010uz, *White:2011yy}). 
Particularly known examples are the subleading terms in soft photon emissions which are referred to as the Low--Burneet--Kroll theorem \cite{Low:1958sn, *Burnett:1967km}.

Although the original proofs of \eqref{gravity} and \eqref{gauge} were performed in four dimensions, the validity in any number of dimensions was soon established \cite{Schwab:2014xua, *Afkhami-Jeddi:2014fia, Broedel:2014fsa,*Bern:2014vva, Zlotnikov:2014sva, *Kalousios:2014uva}. In this work we focus on amplitudes at tree-level so the only comment on loop amplitudes is that single soft subleading terms have also been studied at higher orders in perturbation theory in, e.g., \cite{Bern:2014oka,*He:2014bga,*Cachazo:2014dia, *Bianchi:2014gla}.

In this paper we study two classes of theories that contain scalar particles. The first class consists of theories with neither color nor flavor structure while the second contains theories with a
$U(N)$ color (or flavor) group for the scalars. For both classes we consider single and double soft scalar limits. For the single soft limit we are only concerned with order of the leading term while for the double soft limit we study the actual structure of the leading and subleading soft factors. The leading behavior of soft scalars has been studied for a long time with perhaps the most well-known result being the Adler's zero \cite{Adler:1964um, 
*Susskind:1970gf}, and very recently were studied as a classification tool for scalar theories in \cite{Cheung:2014dqa}. Double soft limits of scalars have also been studied in ${\cal N}=8$ supergravity as a way to explore its moduli space of vacua and the $E_{7(7)}$ structure of the theory \cite{ArkaniHamed:2008gz}.

In the first class of theories we have a special Galileon theory \cite{Dvali:2000hr, *Nicolis:2008in} which we will refer to as sGal \cite{Cheung:2014dqa,Cachazo:2014xea}, the Dirac--Born--Infeld (DBI) theory, and an Einstein--Maxwell--Scalar (EMS) theory. Here we consider theories with a single scalar field, thus for example EMS is the dimensional reduction of Einstein's theory from $(D{+}1)$ to $D$ dimensions.

For theories in the first class we propose that as two scalars, say particles $n+1$ and $n+2$, become soft as $k_{n{+}1}^\mu=\tau p^\mu$, $k_{n{+}2}^\mu=\tau q^\mu$, any $(n+2)$-particle amplitude behaves as
\be\label{soft012}
{\cal M}_{n+2} = (k_{n+1}\cdot k_{n+2})^m (S^{(0)}+S^{(1)}+S^{(2)}){\cal M}_{n} + {\cal O}(\tau^{2m+4} )
\ee
where $m=1,0,-1$ for sGal, DBI and EMS respectively, and
\begin{align}
S^{(0)}&= \frac{1}{4}\sum_{a=1}^n\left(\frac{(k_a\cdot (k_{n+1}-k_{n+2}))^2}{k_a\cdot (k_{n{+}1}+k_{n{+}2})+k_{n{+}1}\cdot k_{n{+}2}}+k_a\cdot (k_{n{+}1}+k_{n{+}2})+k_{n{+}1}\cdot k_{n{+}2}\right),\\
S^{(1)} &= \frac{1}{2}\sum_{a=1}^{n}\frac{k_a\cdot (k_{n+1}-k_{n+2})}{k_a\cdot (k_{n{+}1}+k_{n{+}2})+k_{n{+}1}\cdot k_{n{+}2}}(k_{n+1,\mu}k_{n+2,\nu}J_a^{\mu\nu}),\\
S^{(2)} &= \frac{1}{2}\sum_{a=1}^n\frac{1}{k_a\cdot (k_{n{+}1}+k_{n{+}2})+k_{n{+}1}\cdot k_{n{+}2}}\left((k_{n+1,\mu}k_{n+2,\nu}J_a^{\mu\nu})^2+ (\frac32-2m) (k_{n{+}1}\cdot k_{n{+}2})^2\right).
\end{align}

Here $S^{(0)}$ is a multiplicative operator which has an expansion in $\tau$ starting at ${\cal O}(\tau^1)$, $S^{(1)}$ is a first-order differential operator starting at ${\cal O}(\tau^2)$, and $S^{(2)}$ is a second order differential operator starting at ${\cal O}(\tau^3)$.

Two comments are in order at this point. The first is that in \eqref{soft012} the kinematic invariant $(k_{n+1}\cdot k_{n+2})$ plays the role of a natural ``dimensionful parameter" needed to link amplitudes with different number of particles. This simple dimensional argument leads to universal formulas for $S^{(0)}$ and $S^{(1)}$, i.e., they are theory independent. The second is that the only dependence on the theory under consideration appears in the multiplicative piece of $S^{(2)}$.

As mentioned above, the second class of theories corresponds to those with a $U(N)$ color (or flavor)
structure. In this class we have the non-linear sigma model (NLSM) \cite{Cronin:1967jq, *Weinberg:1966fm, *Weinberg:1968de} and Yang--Mills--Scalar (YMS), which is the dimensional reduction of Yang--Mills theory. The double soft scalar emission for a color-ordered partial amplitude is proposed to be
\be\label{soft01}
{\cal M}(1,2,\ldots ,n,n+1,n+2) = (k_{n+1}\cdot k_{n+2})^m (S^{(0)} + S^{(1)}){\cal M}(1,2,\ldots ,n) + {\cal O}(\tau^{2m+2})
\ee
with $m=0,-1$ for NLSM and YMS respectively, and
\begin{align}\label{colorS01}
S^{(0)} &=\frac{1}{2}\left(\frac{k_n\cdot(k_{n+1}-k_{n+2})+k_{n+1}\cdot k_{n+2}}{k_n\cdot(k_{n+1}+k_{n+2})+k_{n+1}\cdot k_{n+2}} +\frac{k_1\cdot(k_{n+2}-k_{n+1})+k_{n+2}\cdot k_{n+1}}{k_1\cdot(k_{n+2}+k_{n+1})+k_{n+2}\cdot k_{n+1}}\right),\\
S^{(1)} &= \frac{k_{n+1,\mu}k_{n+2,\nu}}{k_n\cdot(k_{n+1}+k_{n+2})+k_{n+1}\cdot k_{n+2}}J_n^{\mu\nu}+\frac{k_{n+2,\mu}k_{n+1,\nu}}{k_1\cdot(k_{n+2}+k_{n+1})+k_{n+2}\cdot k_{n+1}}J_1^{\mu\nu}.
\end{align}
In this formula $S^{(0)}$ starts at ${\cal O}(\tau^0)$ while $S^{(1)}$ starts at order ${\cal O}(\tau)$. Expanding $S^{(0)}$ in \eqref{colorS01} around $\tau=0$ the leading order becomes
\be
S^{(0)} =\frac{1}{2}\left(\frac{k_n\cdot(k_{n+1}-k_{n+2})}{k_n\cdot(k_{n+1}+k_{n+2})} +\frac{k_1\cdot(k_{n+2}-k_{n+1})}{k_1\cdot(k_{n+2}+k_{n+1})}\right) + {\cal O}(\tau).
\ee
This is the famous double soft factor which although finite depends on the relative directions of the soft particles. This structure is the one carrying the information of the non-linearly realized symmetries of the theory which is hinted by Adler's zero in the NLSM. Although the form has been known for many years and tested in the known amplitudes, a proof to all multiplicities in the NLSM was only recently found using BCFW techniques \cite{Kampf:2013vha}. For conventions and details of all the theories we consider here, please refer to \cite{Cachazo:2014xea}.

Last but not least, we find the double soft photon emission theorem in DBI and EMS theory, with a universal leading-order soft factor. Here we record the result: when particles $n{+}1$ and $n{+}2$ are soft photons, an $(n{+}2)$-point amplitude becomes
\be\label{softphoton}
{\cal M}_{n{+}2}=(k_{n{+}1}\cdot k_{n{+}2})^{m{-}1} S^{(0)}{\cal M}_n+{\cal O}(\tau^{2m{+}2})\,,\qquad 
S^{(0)}=\frac14 \sum_{a=1}^n \frac{(k_a\cdot (k_{n{+}1}-k_{n{+}2}))^2}{k_a\cdot (k_{n{+}1}+k_{n{+}2})}\,{\rm Pf} {\cal S}_a\,.
\ee
where $m=0, -1$ for DBI and EMS respectively, and the $4\times4$ anti-symmetric matrix $\mathcal{S}_a$ is
\begin{equation}
\mathcal{S}_a:=\left(
\begin{array}{cccc}
0 & k_{n{+}1} \cdot k_{n{+}2} & k_a^{\perp}\cdot\epsilon_{n{+}1} & -k_{n{+}1}\cdot\epsilon_{n{+}2}\\
-k_{n{+}1}\cdot k_{n{+}2} & 0 & k_{n{+}2}\cdot\epsilon_{n{+}1} & k_a^{\perp'}\cdot\epsilon_{n{+}2}\\
-k_a^{\perp}\cdot \epsilon_{n{+}1} & -k_{n{+}2}\cdot \epsilon_{n{+}1} & 0 & \epsilon_{n{+}1}\cdot\epsilon_{n{+}2}\\
k_{n{+}1}\cdot \epsilon_{n{+}2} & -k_a^{\perp'}\cdot \epsilon_{n{+}2} & -\epsilon_{n{+}1}\cdot\epsilon_{n{+}2} & 0
\end{array}
\right),
\end{equation}
where $(k_a^{\perp})^\mu:=\frac{2 k_{n{+}1} \cdot k_{n{+}2}}{k_a\cdot (k_{n{+}1}{-}k_{n{+}2})}(k_a^\mu-\frac{k_{n{+}1}\cdot k_a}{k_{n{+}1}\cdot k_{n{+}2}}\,k^\mu_{n{+}2})$ and $(k_a^{\perp'})^\mu:=\frac{2 k_{n{+}1} \cdot k_{n{+}2}}{k_a\cdot (k_{n{+}1}{-}k_{n{+}2})}(k^\mu_a-\frac{k_{n{+}2}\cdot k_a}{k_{n{+}1}\cdot k_{n{+}2}}\,k^\mu_{n{+}1})$ are two vectors satisfying $k_a^{\perp}\cdot k_{n{+}1}=0$ and $k_a^{\perp'} \cdot k_{n{+}2}=0$. Gauge invariance of the soft photon factor becomes manifest due to the Pfaffian structure. 

In the rest of the paper we provide various levels of evidence for all the soft theorems proposed here. In Section \ref{sec:check} we list a variety of explicit checks performed with known formulas for the amplitudes. In Section \ref{sec:proof} we use the recently found Cachazo--He--Yuan (CHY) representation~\cite{Cachazo:2013gna, *Cachazo:2013hca, *Cachazo:2013iea} of the theories under study~\cite{Cachazo:2014nsa,Cachazo:2014xea} to give a very simple proof of the leading order terms in $S^{(0)}$ for all the theories mentioned above. A proof for the sub-leading order including the leading part of $S^{(1)}$ is also done using the CHY representation and is presented as supplementary material \cite{webpage}, where we also include the proof for the double soft photon theorem.  We end with discussion on future directions in Section \ref{sec:dis}.

\section{Checks and Proofs for the New Theorems}\label{sec:checkproof}
Here we provide strong evidence for the double soft theorems. We perform non-trivial checks using explicit amplitudes, and use the CHY formula to prove some of the theorems. 

As in the case of soft gluon and graviton emission, amplitudes that enter into our soft theorems are distributions containing momentum-conserving delta functions. For both the checks and proofs, it is useful to write versions of the soft relations for stripped amplitudes. Let us define ${\cal M}=\delta^D(\sum_a k_a) M$ for $n$-point and $(n{+}2)$-point amplitudes, and recall that the two soft momenta are $k_{n{+}1}^\mu=\tau p^\mu$, $k_{n{+}2}^\mu=\tau q^\mu$.

As distributional relations, \eqref{soft012} and \eqref{soft01} are equivalent to relations for all independent distributions when an expansion in $\tau$ is performed, i.e., the momentum-conserving delta functions and its derivatives. One can show that the relations for the derivatives are guaranteed by the relations for the delta functions, or the soft theorems for stripped amplitudes
\begin{align}
[M_{n{+}2}](\tau)&=\left[(\tau^2 p\cdot q)^m(S^{(0)}+S^{(1)}+S^{(2)})\,M_n\right](\tau) + {\cal O}(\tau^{2m{+}4})\,,\\
[M(1,2,\ldots, n{+}1,n{+}2)](\tau)&=\left[(\tau^2 p\cdot q)^m (S^{(0)}+ S^{(1)})\,M(1,2,\ldots ,n)\right](\tau)+ {\cal O}(\tau^{2m{+}2})\,.
\end{align}

In both cases, one first computes LHS and RHS as rational functions of kinematic data (in particular, they are functions of $\tau$), and then evaluate them on the {\it same} kinematics of $(n{+}2)$ momenta that add up to zero. The statement of the double soft emission theorems is that the two are equal at the corresponding orders in $\tau$. It is this form of the soft theorems that we check explicitly and prove in this section.

\subsection{Explicit Checks for the New Theorems}\label{sec:check}

For EMS, DBI and sGal theories, we write down the explicit form of the soft factors:
\begin{align}\label{soft012tau}
S^{(0)}&=\frac {\tau}{4} \sum_{a=1}^n \left(\frac{(k_a\cdot (p-q))^2 }{k_a \cdot (p+q) +\tau p\cdot q}+k_a \cdot (p+q) +\tau p\cdot q\right),\\
S^{(1)}&=\frac{\tau^2}{2} \sum_{a=1}^n \frac{k_a \cdot(p-q) }{k_a\cdot (p+q) + \tau p\cdot q}\,p_\mu q_\nu J^{\mu\nu}_a,\\
S^{(2)}&=\frac{\tau^3}{2} \sum_{a=1}^n \frac{1}{k_a\cdot (p+q) + \tau p\cdot q}\,\left((p_\mu q_\nu J^{\mu\nu}_a)^2+ (\frac32-2m) (p\cdot q)^2\right)\,.
\end{align}
A fact that was not mentioned in Section \ref{sec:intro} but crucial in the definition of $S^{(1)}$ and $S^{(2)}$ is that the operator $J_a$ does not annihilate the prefactor $(k_a\cdot (p{+}q) + \tau p\cdot q)^{-1}$ and so the above expressions literally mean that the prefactor comes in front of $J_a$.

We computed pure scalar amplitudes in DBI theory analytically up to ten points, and checked  the theorem at ${\cal O}(\tau), {\cal O}(\tau^2), {\cal O}(\tau^3)$ for $n=4,6$, and at ${\cal O}(\tau), {\cal O}(\tau^2)$ for $n=8$. We have also computed scalar amplitudes in EMS and sGal theory explicitly up to eight points, and confirmed the complete soft scalar theorems for $n=4,6$.

In addition, we performed two important checks in DBI theory for amplitudes with external photons. (i) For the subleading soft scalar factors, the angular momentum of a photon has an additional spin part, $J_{a, {\rm photon}}=J_{a,{\rm scalar}}+J_{a, {\rm spin}}$, which acts on the polarization vectors as $(J_{a, {\rm spin}}^{\mu \nu}\, \epsilon_a)^\beta=(\eta^{\nu \beta}\delta^{\mu}_{\sigma}-\eta^{\mu \beta}\delta^{\nu}_{\sigma})\,\epsilon_a^\sigma$. We checked the leading and sub-leading theorems for the four-scalar-two-photon DBI amplitude, where in $S^{(1)}$, $J_{\rm photon}$ is used for acting on the photons. (ii) We also checked the soft photon theorem for six-photon amplitudes in DBI.

For partial amplitudes in YMS and NLSM theories, the soft operators read
\begin{align}\label{soft01tau}
S^{(0)}&=\frac 1 2 \left(\frac{k_n\cdot(p{-}q) +\tau p\cdot q}{k_n\cdot(p{+}q) +\tau p\cdot q}+\frac{k_1\cdot(q{-}p) +\tau q\cdot p}{k_1\cdot(q{+}p) +\tau q\cdot p}\right)\,,\\
S^{(1)}&=\tau\left(
\frac{1}{k_n\cdot(p{+}q) +\tau p\cdot q}p_\mu q_\nu J_n^{\mu\nu}+
\frac{1}{k_1\cdot(q{+}p) +\tau q\cdot p}q_\mu p_\nu J_1^{\mu \nu}\right)\,.
\end{align}
We computed NLSM amplitudes up to ten points, and checked the theorem at ${\cal O}(\tau^0), {\cal O}(\tau^1)$ for $n=4,6,8$. We also confirmed the theorem for six-scalar amplitudes in YMS theory.
\subsection{CHY Representation and Soft Limits}\label{sec:limits}

Before proceeding to the proofs, let us first review the CHY representation for amplitudes in these theories \cite{Cachazo:2013hca}. It is given by an integral over the moduli space of $n$-punctured Riemann spheres, with the locations specified by holomorphic variables $\sigma$'s:
\begin{equation}\label{CHY}
M_n=\int \prod_{a=1}^n{}' d\sigma_a\,\prod_{a=1}^n{}'\delta(f_a)\,I_n(\{\sigma, k,\ldots\})=:\int d\mu_n\,I_n,
\end{equation}
where the delta functions of $f_a:=\sum_{b\neq a}\frac{k_a\cdot k_b}{\sigma_a-\sigma_b}$ impose the so-called scattering equations \cite{Cachazo:2013gna}, and the integrand $I_n$ is some rational function that depends on the theory under consideration. The primes in the products denote redundancies in both the variables and the equations: for each product, one has to exclude three labels and compensate by a factor, e.g., $\prod'{}_a:= \sigma_{i,j}\,\sigma_{j,k}\,\sigma_{k,i}\prod_{a\neq i,j,k}$ (here $\sigma_{a,b}:=\sigma_{a}-\sigma_{b}$), and the result is independent of the choice.

For scalar amplitudes of the theories we study in this paper, the integrand $I_n$ is a combination of three basic building blocks~\cite{Cachazo:2014nsa}: (i) the Parke--Taylor factor for partial amplitudes with a given ordering, say with the canonical ordering $(1,2,\ldots,n)$
\begin{equation}
C(1,2,\ldots,n):=\frac{1}{(\sigma_1-\sigma_2)(\sigma_2-\sigma_3)\,\cdots\,(\sigma_n-\sigma_1)},
\end{equation}
(ii) the Pfaffian ${\rm Pf}X_n$, and (iii) the reduced Pfaffian ${\rm Pf}'A_n$, where the two $n\times n$ anti-symmetric matrices are defined by specifying their entries as
\begin{equation}\label{defXA}
(X_n)_{ab}:=\frac{1}{\sigma_a-\sigma_b}\,(1-\delta_{ab})\,,\quad
(A_n)_{ab}:=\frac{k_a\cdot k_b}{\sigma_a-\sigma_b}\,(1-\delta_{ab})\,.
\end{equation}
Since the matrix $A_n$ has co-rank $2$ on the support of the scattering equations $f_a=0$, we define the invariant quantity ${\rm Pf}'A_n:=\frac{(-1)^{i+j}}{\sigma_i-\sigma_j}{\rm Pf}|A_n |^{i,j}_{i,j}$, where the minor $|A_n|^{i,j}_{i,j}$ is obtained by deleting the $i^{\rm th}$ and $j^{\rm th}$ rows and columns. With these building blocks, the integrands for scalar amplitudes of the theories we study are constructed as follows
\begin{align}
&I_n^{\rm sGal}:=({\rm Pf}'A_n)^4\,,\quad
I_n^{\rm DBI}:={\rm Pf}X_n\,({\rm Pf}'A_n)^3\,,\quad
I_n^{\rm EMS}:=({\rm Pf}X_n)^2\,({\rm Pf}'A_n)^2\,,\\
&I_n^{\rm NLSM}(1,2,\ldots,n):=C(1,2,\ldots,n)\,({\rm Pf}'A_n)^2\,,\quad
I_n^{\rm YMS}(1,2,\ldots,n):=C(1,2,\ldots,n)\,{\rm Pf}X_n\,{\rm Pf}'A_n\,.\nonumber
\end{align}

The integrals in \eqref{CHY} are localized on the $\sigma$ solutions to the scattering equations. With this formula the leading order of the above amplitudes under single and double soft emissions can be easily extracted by studying the behavior of the solutions for $\sigma$'s first.

For single soft emission, 
all the solutions are non-degenerate in the sense that $(\sigma_a-\sigma_b)\sim \tau^0$ for any $a,b$. For double soft emission, 
however, apart from such non-degenerate solutions there exists a unique {\it degenerate} solution in which $(\sigma_{n+1}-\sigma_{n+2})\sim\tau$. Given these we obtain the leading scaling in $\tau$ of $d\mu$ and of the building blocks on the solutions, which are summarized in Table \ref{table1} 
(we use ``d'' to denote the degenerate solution and ``n'' the non-degenerate ones).
\begin{table}[h]\centering
\caption{Leading Scaling of the Building Blocks}\label{table1}
\begin{tabular}{c||c||c|c}
\hline
& single soft & double soft (n) & double soft (d)\\
\hline\hline
$d\mu$ & $\tau^{-1}$ & $\tau^{-2}$ & $\tau^{-1}$\\
\hline
$C$ & $\tau^0$ & $\tau^{0}$ & $\tau^{-1}$\\
\hline
${\rm Pf}X$ & $\tau^0$ & $\tau^{0}$ & $\tau^{-1}$\\
\hline
${\rm Pf}'A$ & $\tau^1$ & $\tau^{2}$ & $\tau^{1}$\\
\hline
\end{tabular}
\end{table}
As a consequence, the leading scaling of the formulas are summarized in Table \ref{table2}. Now it becomes obvious that double-soft behavior of these theories are special in that the contributions from non-degenerate solutions are suppressed: only the degenerate solution contributes to the first three and two orders for the two classes of theories respectively, which are exactly the orders we considered in the soft theorems.
\begin{table}[h]\centering
\caption{Leading Scaling of the Formulas}\label{table2}
\begin{tabular}{c||c||c|c}
\hline
& single soft & double soft (n) & double soft (d)\\
\hline\hline
sGal & $\tau^{3}$ & $\tau^{6}$ & $\tau^{3}$\\
\hline
DBI & $\tau^{2}$ & $\tau^{4}$ & $\tau^{1}$\\
\hline
EMS & $\tau^{1}$ & $\tau^{2}$ & $\tau^{-1}$\\
\hline\hline
NLSM & $\tau^{1}$ & $\tau^{2}$ & $\tau^{0}$\\
\hline
YMS & $\tau^{0}$ & $\tau^{0}$ & $\tau^{-2}$\\
\hline
\end{tabular}
\end{table}

\subsection{Proof of the Leading-Order Double Soft Theorems}\label{sec:proof}

Here we present the proof for the double soft scalar theorems at the leading order (see \cite{webpage} for the proof at sub-leading order, and that for the double photon theorem.). As discussed before it suffices to consider the degenerate solution only, for which we make the following transformation (we choose not to delete $d\sigma_{n+1}d\sigma_{n+2}\delta(f_{n{+}1})\delta(f_{n{+}2})$):
\ba
&&\sigma_{n+1}=\rho-\frac{\xi}{2}\,,\quad
\sigma_{n+2}=\rho+\frac{\xi}{2}\,,\nl
&& d\sigma_{n+1}\,d\sigma_{n+2}\,\delta(f_{n+1})\,\delta(f_{n+2})
=-2\,d\rho\,d\xi\,\delta(f_{n+1}+f_{n+2})\,\delta(f_{n+1}-f_{n+2}).
\ea
The basic idea here is to localize the $\xi$-integral by $\delta(f_{n{+}1}-f_{n{+}2})$, and regard the $\rho$-integral as a contour integration whose contour wraps the zeros of the equation $f_{n{+}1}+f_{n{+}2}=0$. Since $\xi=\sigma_{n{+}2}-\sigma_{n{+}1}\sim\tau$, we can further expand it as $\xi=\tau\,\xi_1+\mathcal{O}(\tau^2)$. Starting from the formula for $M_{n+2}$ this leads us to
\begin{equation}\label{highexpandpre}
M_{n+2}=-\int d\mu_{n}\,\oint\frac{d\rho}{2\pi i}\,
\frac{1}{\sum_{a=1}^n\frac{k_a\cdot(p+q)}{\rho-\sigma_a}}\,
\frac{\xi_1^2}{\tau\,p\cdot q}\,I_{n+2}+\text{(sub-leading)}\,,
\end{equation}
where the $\rho$-contour is now specified by $\sum_{a=1}^n\frac{k_a\cdot(p+q)}{\rho-\sigma_a}=0$, and $\xi_1$ is evaluated on its unique solution $
\xi_1^{-1}=\frac{1}{2p\cdot q}\sum_{a=1}^{n}\frac{k_a\cdot(p-q)}{\rho-\sigma_a}$.

To derive the leading-order soft theorems from \eqref{highexpandpre}, we need to expand $I_n$ with respect to $\tau$. We first look at each building block, which behaves as
\begin{align}
C(1,2,\ldots,n+2)&=C(1,2,\ldots,n)\,\frac{\sigma_n-\sigma_1}{(\sigma_n-\rho)\,(-\tau\,\xi_1)\,(\rho-\sigma_1)}+\mathcal{O}(\tau^0)\,,\\
{\rm Pf}X_{n+2}&=-\frac{1}{\tau\,\xi_1}\,{\rm Pf}X_n+\mathcal{O}(\tau^0)\,,\\
{\rm Pf}'A_{n+2}&=-\frac{\tau\,p\cdot q}{\xi_1}\,{\rm Pf}'A_n+\mathcal{O}(\tau^2)\,.
\end{align}
Combining these we obtain for scalar amplitudes in sGal, DBI and EMS (with $m=1,0,-1$)
\begin{equation}
\begin{split}
M_{n+2}&=-\int d\mu_{n}\,I_n\,\oint\frac{d\rho}{2\pi i}\,
\frac{1}{\sum_{a=1}^n\frac{k_a\cdot(p+q)}{\rho-\sigma_a}}\,
\frac{\xi_1^2}{\tau\,p\cdot q}\,\left(\frac{1}{\tau\,\xi_1}\right)^{1{-}m}\left(\frac{\tau\,p\cdot q}{\xi_1}\right)^{m{+}3}+\mathcal{O}(\tau^{2m{+}2})\\
&=-\int d\mu_{n}\,I_n\,\frac{\tau\,(\tau^2\,p\cdot q)^m}{4}\oint\frac{d\rho}{2\pi i}\,
\frac{\left(\sum_{a=1}^n\frac{k_a\cdot(p-q)}{\rho-\sigma_a}\right)^2}{\sum_{a=1}^n\frac{k_a\cdot(p+q)}{\rho-\sigma_a}}+\mathcal{O}(\tau^{2m{+}2})\,.
\end{split}
\end{equation}
Now we perform the $\rho$-integral by deforming the contour and use a residue theorem. Although there appears to be a simple pole at $\rho=\infty$, it is eliminated by an additional zero in the numerator due to momentum conservation. Thus we only encounter simple poles at $\rho=\sigma_a$ ($a=1,\ldots,n$), and the final result is
\begin{equation}
M_{n+2}=(\tau^2\,p\cdot q)^m\left(\frac{\tau}{4}\sum_{a=1}^n
\frac{\left(k_a\cdot(p-q)\right)^2}{k_a\cdot(p+q)}\right)M_n+\mathcal{O}(\tau^{2m{+}2})\,.
\end{equation}
Note that the soft operator in the bracket agrees with $S^{(0)}$ in \eqref{soft012tau} at ${\cal O}(\tau)$, because the additional piece in $S^{(0)}$ is of higher order by momentum conservation, $\sum_{a=1}^n k_a\cdot (p+q)=-2\tau (p.q)^2$. This concludes our proof for the leading order soft theorem in these theories.

Similarly, for scalar partial amplitudes in NLSM and YMS (with $m=0,-1$) we obtain
\begin{align}
&M(1,2,\ldots,n+2)\nonumber\\
&=-\int d\mu_{n}\,I_n\,\oint\frac{d\rho}{2\pi i}\,
\frac{\xi_1^2/p\cdot q}{\tau\sum_{a=1}^n\frac{k_a\cdot(p+q)}{\rho-\sigma_a}}\,
\frac{\sigma_n{-}\sigma_1}{(\sigma_n{-}\rho)(-\tau\,\xi_1)(\rho{-}\sigma_1)}\left(\frac{1}{\tau\xi_1}\right)^{-m}\left(\frac{\tau\,p\cdot q}{\xi_1}\right)^{2+m}+\mathcal{O}(\tau^{2m{+}1})\nonumber\\
&=\int d\mu_{n}\,I_n\,\frac{(\tau^2 p\cdot q)^m}{2} \oint\frac{d\rho}{2\pi i}\,
\frac{\sum_{a=1}^n\frac{k_a\cdot(p-q)}{\rho-\sigma_a}}{\sum_{a=1}^n\frac{k_a\cdot(p+q)}{\rho-\sigma_a}}\,\frac{\sigma_n{-}\sigma_1}{(\sigma_n{-}\rho)\,(\rho{-}\sigma_1)}+\mathcal{O}(\tau^{2m{+}1})\,,
\end{align}
Obviously there is no pole at $\rho=\infty$, and we pick up two poles at $\rho=\sigma_1$ and $\rho=\sigma_n$, which lead to the correct leading order soft theorem in these theories,
\begin{equation}
M(1,2,\ldots,n+2)=\frac{(\tau^2 p\cdot q)^m}{2}\left(\frac{k_n\cdot(p-q)}{k_n\cdot(p+q)}+\frac{k_1\cdot(q-p)}{k_1\cdot(q+p)}\right)M(1,2,\ldots,n)+\mathcal{O}(\tau^{2m{+}1})\,.
\end{equation}

\section{Discussion}\label{sec:dis}

Although not mentioned in Section \ref{sec:intro}, the original motivation to search for subleading soft theorems in graviton emission in \cite{Cachazo:2014fwa} was the connection among soft theorems and Ward identities of BMS symmetries~\cite{Bondi:1962px, *Sachs:1962wk, Barnich:2009se, *Barnich:2011ct, *Barnich:2011mi} at null infinity \cite{Strominger:2013jfa, *He:2014laa}. Very recently, a Kac--Moody structure was found for four dimensional Yang--Mills also at null infinity and two consecutive soft limits play an important role \cite{He:2015zea}. It is very tempting to suggest that the double soft theorems for scalars and the ones for photons could also have an interpretation as hidden symmetries. It is well known that in theories where single soft scalar limits vanish, double soft scalar factors carry information about non-linearly realized symmetries. It would be interesting to carefully explore the meaning of each of the soft theorems we found.

As explained in Section \ref{sec:limits}, the CHY integrand evaluated on different solutions to the scattering equations behaves in different ways, as summarized in Table \ref{table2}. There was a single degenerate solution $({\rm d})$ while all the others were non-degenerate $({\rm n})$.  Denoting the leading double soft behavior of the integrand on a given solution as ${\cal O}(\tau^{\alpha_{\rm d}})$ or ${\cal O}(\tau^{\alpha_{\rm n}})$, one finds that the presence of soft factors is related to the difference
\be
\Delta \equiv \alpha_{\rm n}-\alpha_{\rm d}.
\ee
We have seen that in all theories where $\Delta = 3$ one finds the presence of $S^{(0)}, S^{(1)}$, and $S^{(2)}$ while in theories with $\Delta = 2$ only $S^{(0)}, S^{(1)}$ are present. One can also compute $\Delta$ for the emission of other kind of particles, e.g., photons in DBI and EMS theory. In \eqref{softphoton} we provided a very compact formula for $S^{(0)}$ of double soft photon emission and one might wonder whether there exists sub-leading soft factors. In both theories $\Delta = 1$ which suggests that there is no other universal soft factor. Finding a clear explanation of the relation between $\Delta$ and the presence of soft factors is clearly one of the directions of future research.

A related question is the derivation of the new soft theorems using BCFW techniques as was done originally in \cite{Cachazo:2014fwa}. A clear first problem is that BCFW techniques present technical challenges when applied to scalar theories if no extra symmetries (such as supersymmetry) are present. However, we believe that $\Delta$ might somehow control the behavior for infinity momenta of some generalization of BCFW appropriate for these theories. Very recently a complete analysis of applications of BCFW-like techniques was carried out in \cite{Cheung:2015cba} and perhaps their approach can shed some light on this issue.

We proved some of the soft theorems using the CHY representation, which appears to be particularly powerful for studying soft behavior of amplitudes. An important point to mention is that some of the CHY formulas here are still conjectural. This is why the checks done on explicitly known amplitudes are relevant. The still conjectural CHY representations have passed many non-trivial tests such as factorization. Therefore our proofs using them are very strong evidence for the validity of the leading and sub-leading soft theorems.

Finally, let us mention an interesting analogy between the single soft theorem of a graviton (gluon) and the double soft theorem of theories of the first (second) class. Consider the single graviton soft factors and replace the polarization tensor $\epsilon_{\mu\nu}$ by the product of two polarization vectors, i.e., $\epsilon_\mu\epsilon_\nu$. Then
\be
S^{(0)}_{\rm gravity} = \sum_{a=1}^n\frac{(\epsilon_{\mu}k^\mu_a)^2}{k_{n+1}\cdot k_a}, ~~
S^{(1)}_{\rm gravity} = \sum_{a=1}^n\frac{(\epsilon_{\mu}k^\mu_a) (\epsilon_\nu k_{n+1,\rho}J^{\rho\nu}_a)}{k_{n+1}\cdot k_a}, ~~ S^{(2)}_{\rm gravity} = \frac{1}{2}\sum_{a=1}^n\frac{(\epsilon_{\mu}k_{n+1,\rho}J^{\rho\mu}_a)^2}{k_{n+1}\cdot k_a}.
\ee
Comparing this form with the double soft factors for theories in the first class \eqref{soft012} it is easy to see that by identifying
\be
\epsilon_\mu \mapsto k_{n+1,\mu} - k_{n+2,\mu} \quad {\rm and} \quad k_{n+1,\,{\rm soft\,graviton}}^\mu \mapsto k_{n+1}^\mu + k_{n+2}^\mu
\ee
one finds a striking similarity. The same analogy exists for theories of the second class and the single soft gluon emission. It would be interesting to sharpen this connection.

{\it Acknowledgements:}
Research at Perimeter Institute is supported by the Government of Canada through Industry Canada and by the Province of Ontario through the Ministry of Research \& Innovation.

\bibliographystyle{apsrev4-1}
\bibliography{double_soft}

\end{document}